\documentclass[preprint,prd,noshowpacs,nofootinbib]{revtex4}

\usepackage{amssymb}
\usepackage{amsmath}
\usepackage{graphicx}  % needed for figures
\usepackage{hyperref}

\usepackage{color}

\usepackage{lipsum}

\usepackage[normalem]{ulem}

\usepackage[dvipsnames]{xcolor}

\usepackage{setspace}

\newcommand{\mathsym}[1]{{}}
\newcommand{\unicode}[1]{{}}

\usepackage{tablefootnote}

\begin{document}

\title{Gravito-optics and intensity correlations for binary inspiral signal detections}

\author{Preston Jones}
\email{Preston.Jones1@erau.edu}
\author{Alexander Barrett}
\author{Justin Carpenter}
\author{Andri Gretarsson}
%\email{Andri.Gretarsson@erau.edu}
\author{Ellie Gretarsson}
%\email{EJesse@email.arizona.edu}
\author{Brennan Hughey}
%\email{Brennan.Hughey@erau.edu}
\author{Darrel Smith}
%\email{Smith@erau.edu}
\author{Michele Zanolin}
%\email{Michele.Zanolin@erau.edu}
\affiliation{Embry Riddle Aeronautical University, Prescott, Arizona 86301, USA }
\author{Douglas Singleton}
\email{dougs@mail.fresnostate.edu}
\affiliation{California State University Fresno, Fresno, California 93740, USA}

\date{\today}

\begin{abstract}
We examine the correlation functions associated with intensity interferometry and gravito-optics of gravitational wave signals from compact binary coalescences. Previous theoretical studies of the gravito-optics of gravitational waves has concentrated on the characterization of both the classical and the non-classical properties of signals from cosmological sources in the early Universe. These previous works assume a periodic signal similar to the signals studied widely in optics and quantum optics and do not apply to transient signals. We develop the gravito-optics of intensity correlations for descriptions of the detection of transient signals from compact binary coalescences and apply these methods to calculate the two-point intensity correlations for the gravitational wave discovery. We also discuss the necessary theoretical work required for the description of the quantum gravito-optics of intensity correlations in the detection of signals from binary inspirals.
\end{abstract}

\maketitle

\section{Intensity correlations for compact binary coalescences}\label{Introduction}

\begin{figure}[htp!]
 \centering
\includegraphics[width=90mm]{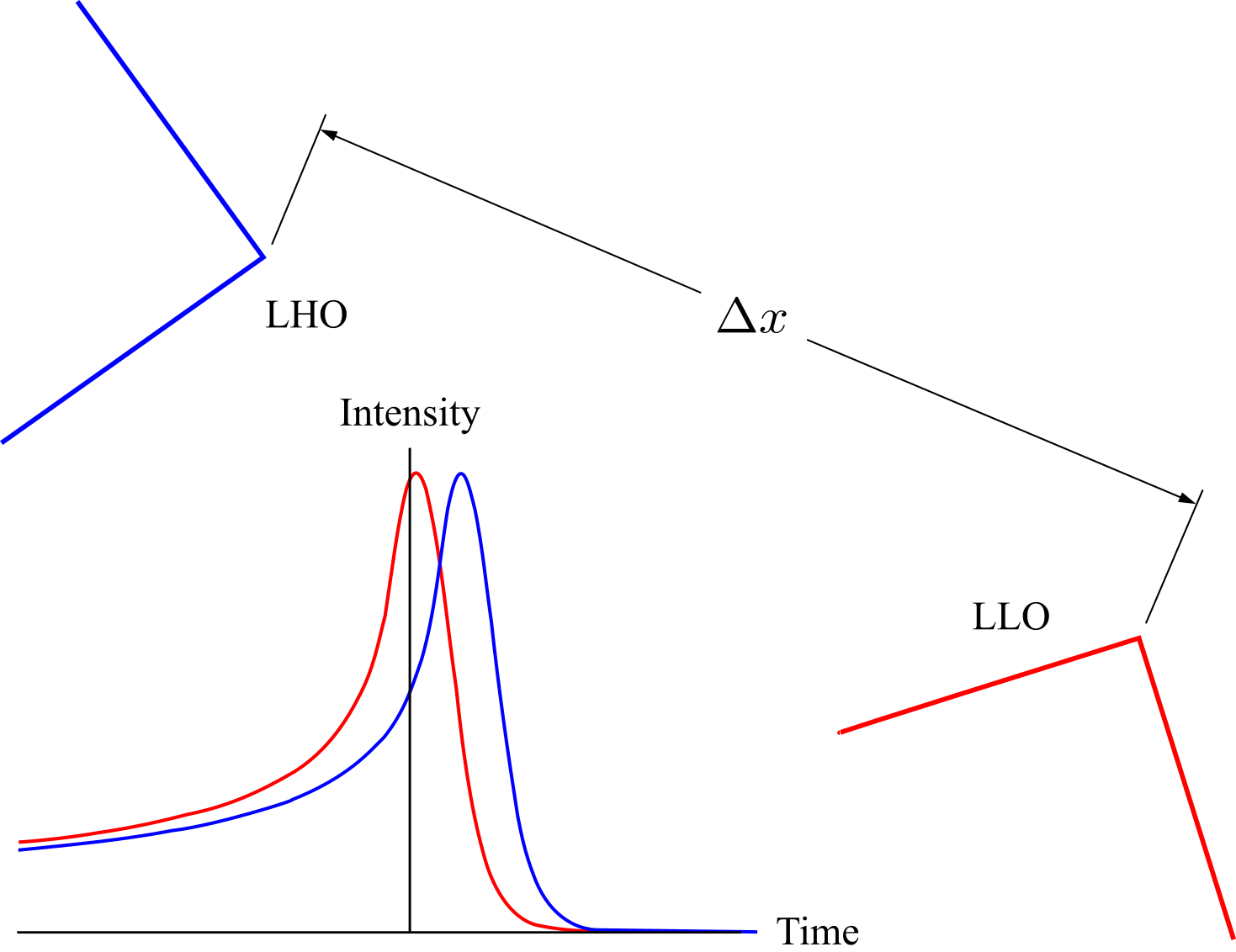}
 \caption{GW observatories that operate concurrently form an intensity or HBT interferometer. \href{https://www.ligo.org/scientists/GW100916/GW100916-geometry.html}{Spatially separated observatories} \cite{LIGO}, $ \Delta x \sim  3,000 ~ \rm{km} $, exhibit two-point detection of GW signals from CBC with lag time $\tau$ between detections. Intensity correlations for GW signals are calculated as a function of this time lag between signal detections.}
\label{fig}
\end{figure}

The concurrent operation of independent and physically separated detectors in optics and particle physics form two-point intensity interferometers or Hanbury Brown and Twiss (HBT) interferometers. Since existing gravitational wave (GW) observatories can be and are operated concurrently these observatories can also be used as HBT intensity interferometers as illustrated in Fig. \ref{fig}. While HBT interferometry for primordial GW sources has been well studied the application of HBT interferometry to Binary inspirals or compact binary coalescences (CBC) is not well developed \cite{Unnikrishnan15}. Our goal is to expand on the well established applications of intensity correlations in optics to gravitational radiation or gravito-optics. Specifically to address the unique challenges of calculating two-point intensity correlations for transient signal detections from CBC that are currently observable. This work will build on the applications of intensity correlation functions for studies of steady signals from early work on cosmological and primordial sources of gravitational radiation to the gravito-optics of CBC.

We consider first and second order correlations \cite{Glauber63,Glauber63_1,Sudarshan63}, that are centrally important for the characterization of optical signal detections, for GW detections and for the hypothetical graviton. Correlation functions are widely used to characterize the classical and non-classical features for signal detection using HBT interferometry in astrophysics \cite{Brown56,Malvimat14} and particle physics \cite{Baym98,Baym99}. We develop methods for calculation of these correlation functions for the transient GW signals from CBC. These correlations are the foundation of quantum optics \cite{Mandel95,Loudon,Fox} and are equally important in particle physics \cite{Baym98,Baym99}. An important distinction between GW HBT interferometry and previous applications of HBT interferometry, is that GW detectors, independently and directly, measure the amplitude of the GW. In traditional HBT interferometry, used in astrophysics or particle physics, it is the intensity ({\it i.e.} the square of the amplitude) which is measured. For this reason traditional HBT interferometry is often called intensity interferometry. Applying HBT interferometry to the detection of GWs from CBC leads to important differences from traditional HBT interferometry which we address in this paper after a general discussion of intensity correlations.

%-----------------------------Figure Start------------------------------
% encapsulated postscript
%-----------------------------Figure End--------------------------------

Several recent papers \cite{Giovannini11,Giovannini17,Giovannini17b,Giovannini19a,Kanno18,Giovannini19,Kanno19} have discussed the possibility of applying HBT interferometry to the characterization of cosmological gravitational waves (CGW) \cite{grischuk,grischuk1,grischuk2}. These works suggest that one could potentially detect the non-classicality of CGW in squeezed, coherent\footnote{Following \cite{Baym98,Baym99,Loudon,Fox} in our discussion of measurements of correlations, coherent radiation refers to ``stable waves''  \cite{Loudon} and specifically a constant temporal second order ``degree of coherence'', $g^{(2)} \left( \tau \right) = 1$.}  \cite{Glauber63,Baym98,Brown56,Loudon,Fox} states by measuring their correlation functions with {\it future} space based detectors. Binary inspirals are not expected to produce squeezed states \cite{Unnikrishnan15,Giovannini19,Kanno19} and these discussions of non-classicality {\it do not} apply to GWs from CBC. However, non-classical contributions to the detection of GW signals from CBC could be meaningful with third generation detectors which is discussed in Section \ref{multi-graviton}. This is due to the fortuitously low efficiency of GW detectors.  Fortuitous in the sense that the low efficiency would be associated with a comparable low number expectation value for detected gravitons. Low expectation values for graviton number could increase the relative contributions of non-classical phenomena. We more generally discuss many of the possibilities for identification of non-classicality in detections of GW signals from CBC and show that these contributions might be significant for {\it future} detectors in Sections \ref{efficiency} and \ref{multi-graviton}.

One interesting recent paper \cite{Parikh21} examines the possibility of non-classical contributions to single detector processes for GW signals. These possibilities would be associated with single graviton responses which, as described by Dyson \cite{Dyson13}, are generally considered to be excluded based on fundamental physical arguments  \cite{Dyson13,Kanno21}. In contrast we consider the possibility of multi-graviton contributions to non-classicality similar to two-point intensity detectors in quantum-optics and particle physics. We also discuss the work needed to provide a theoretical framework for appropriate criteria for identification of non-classicality in GW signal detection. This section also includes a discussion of the need for an order of magnitude improvement over existing gravitational wave detector sensitivities if experimental verification of non-classicality is to be possible. Calculations of correlation functions for two-point detections of GW signals from CBC are not the same as previous applications to optics, particle physics, or CGW. Binary inspiral signals are both transient and amplitude modulated. With few exceptions \cite{Lebreton13,Ragy13} previous calculations of correlation functions have been for persistent and periodic signals. Calculation of correlation functions for transient and amplitude modulated GW signals from CBC are developed in Section \ref{binary}. 

\section{Intensity correlation functions}\label{HBT}

First and second order correlation functions are centrally important in modern studies of optics, as well as radio astronomy and particle physics \cite{Glauber63,Glauber63_1,Sudarshan63,Mandel95,Loudon,Fox,Malvimat14,Baym98,Baym99}. These correlations have also been employed in theoretical studies of primordial GWs \cite{Giovannini11,Giovannini17,Giovannini17b,Giovannini19a,Kanno18,Giovannini19,Kanno19} where the signals are expected to share the periodicity commonly found in optical signals. However, the signals from CBC are transient and this previous work does not directly apply to calculations of correlation functions for these signals. Applications of correlation functions to studies of GWs from CBC requires the development of unique techniques for calculation of the correlations. To provide a context for our discussion of these new techniques we present a brief description of the fundamentals of first and second order correlations in this Section.

A generic analysis of the HBT interferometry and correlation functions can be based on the classical, normalized first and second order correlation functions \cite{Glauber63,Loudon,Fox}. The first order correlation functions (also called amplitude correlations) for general classical field amplitudes, $\psi$, are defined as

\begin{equation}
g^{\left( 1 \right)} \left( \tau  \right) = \frac{{\left\langle { \psi^* \left( {t + \tau } \right) \psi\left( t \right)} \right\rangle }}{  \sqrt{\left\langle {\psi^* \left( t + \tau \right)\psi\left( t + \tau\right)} \right\rangle } \sqrt{\left\langle {\psi^* \left( t \right)\psi\left( t \right)} \right\rangle }} ,
\label{1correlationfunction}
\end{equation}
~\\

\noindent where $\tau$ is a lag time between spatially separated independent detectors. It is conventional to assume a constant weighting function for the integral time averages,
$\left\langle {{\psi ^*}\left( {t + \tau } \right)\psi \left( t \right)} \right\rangle  = \frac{1}{T}\int_T {{\psi ^*}\left( {t + \tau } \right)\psi \left( t \right)} dt$. However, this time average is not meaningful for transient signals and a more suitable weighting function will be developed in Section \ref{binary}. In HBT intererometry the detectors are separated by a baseline, the distance between physically separated detectors, which leads to a difference in the detector response time $  \tau $ \cite{Abbott20}. The associated second order correlation functions (also called intensity correlations) for classical field amplitudes, $\psi$, are

\begin{equation}
g^{\left( 2 \right)} \left( \tau  \right) = \frac{{\left\langle {\psi^* \left( t \right)\psi^* \left( {t + \tau } \right)\psi\left( {t + \tau } \right)\psi\left( t \right)} \right\rangle }}{{\left\langle {\psi^* \left( t \right)\psi\left( t \right)} \right\rangle }{\left\langle {\psi^* \left( t + \tau \right)\psi\left( t + \tau\right)} \right\rangle }} .
\label{correlationfunction}
\end{equation}

\noindent These second order correlations can be expressed in terms of the first order correlation for an incoherent source as \cite{Fox}, 
$g^{\left( 2 \right)} \left( \tau  \right) = 1 + \left| {g^{\left( 1 \right)} \left( \tau  \right)} \right|^2$. Classical radiation has values of $g^{\left( 2 \right)} \left( 0 \right) \ge 1$, which are Poissonian correlations (for the equal sign) or super-Poissonian correlations (for the greater sign). Non-classical radiation is associated with squeezed states and anti-bunching \cite{Fox,Strekalov17} and has second order correlations which are sub-Poissonian with $g^{\left( 2 \right)} \left( 0 \right) < 1$. While the squeezed states of non-classical radiation are important in quantum optics these states are not produced in CBC and are not relevant to the present work as discussed in the introduction Section \ref{Introduction}.

\subsection{First and second order correlations for GWs }\label{12glauber}
 
In this subsection we translate the above general discussion of intensity correlations to the case of GWs where the generic amplitude $\psi (\tau)$ is replaced by the rank-2 tensor $h_{\alpha \beta} (u)$. This tensor, $h_{\alpha \beta}$ is the deviation from the flat spacetime, $\eta_{\alpha \beta}$ ({\it i.e.} $g_{\alpha \beta} (u) = \eta _{\alpha \beta} + h_{\alpha \beta} (u)$) and is a function of the light cone coordinate $u=t-z$, where $c=1$. The plane wave expansion of $h_{\alpha \beta}$ in the transverse traceless gauge is \cite{Allen96,Renzini18,Isi18},

\begin{equation}
h_{\alpha \beta} \left(u\right)=\sum_A \int_{-\infty}^\infty dk\ \int_{S^2} 
d\hat\Omega\ h_A(k,\hat \Omega)\ e^{iku}\ 
e_{\alpha \beta}^A( \hat \Omega),
\label{strain}
\end{equation}

\noindent where $\hat \Omega$ is the direction normal to the surface $S^2$ and $z = \vec x \cdot \hat \Omega $. The polarization vectors, $e_{\alpha \beta}^A( \hat \Omega)$, are plus and cross states, $A=+,\times$. Taking a single polarization $e_{\alpha \beta}^A( \hat \Omega) \to e_{\alpha \beta}( \hat \Omega)$,

\begin{equation}
h_{\alpha \beta} \left(u\right)=\int_{-\infty}^\infty dk\ \int_{S^2} 
d\hat\Omega\ h(k,\hat \Omega)\ e^{iku} e_{\alpha \beta}( \hat \Omega).
\label{strain2}
\end{equation}

\noindent The detection period for GW signals from CBC typically includes multiple cycles and the frequency can vary over this period. In general, calculation of correlation functions \eqref{1correlationfunction} and \eqref{correlationfunction} would include the frequency variations from \eqref{strain2}. In order to continue our general discussion of the application of first and second order correlations to GW detections we assume a signal far from the source, take the propagation direction as $\hat z$ ({\it i.e.} $h(k,\hat \Omega) \to h(k,\hat z)$) and all the nonzero tensor components in the plane wave expansion \eqref{strain2} are the same, $h_{\alpha \beta} \left(u\right) \to h \left(u\right)$,

\begin{equation}
h\left( u \right) = \int_{ - \infty }^\infty  d k{\kern 1pt} h(k)\;e^{iku} \;,
\label{strain3}
\end{equation}

\noindent where $h(k)$ is the complex-valued Fourier amplitude of wave number $k$ in the plane wave expansion \cite{Isi18}. With this translation, the generic first order correlation of \eqref{1correlationfunction} can be specialized to a GW via \eqref{strain3} with the form

\begin{equation}
g^{\left( 1 \right)} \left( \tau  \right) = \frac{{\left\langle { h^* \left( {t + \tau } \right) h\left( t \right)} \right\rangle }}{  \sqrt{\left\langle {h^* \left( t + \tau \right) h \left( t + \tau\right)} \right\rangle } \sqrt{\left\langle {h^* \left( t \right) h \left( t \right)} \right\rangle }} ,
\label{hCorrelation}
\end{equation}

\noindent where $\tau$ is the delay time between detectors. Similarly, the generic second order correlation function from \eqref{correlationfunction} can be translated, via \eqref{strain3}, into the second order correlation function for GWs

\begin{equation}
g^{\left( 2 \right)} \left( \tau  \right) = \frac{{\left\langle {h^* \left( t \right)h^* \left( {t + \tau } \right)h\left( {t + \tau } \right)h\left( t \right)} \right\rangle }}{{\left\langle {h^* \left( t \right)h\left( t \right)} \right\rangle \left\langle {h^* \left( {t + \tau } \right)h\left( {t + \tau } \right)} \right\rangle }} .
\label{2Correlation}
\end{equation}

\subsection{Gravitational radiation and degrees of coherence}\label{gwCoherence}
 
For electromagnetic radiation one way to distinguish classical radiation from radiation with a non-classical character is through  $g^{\left( 2 \right)} \left( \tau  \right)$ \cite{Fox}.  Classical radiation has $g^{(2)} \left( 0 \right) \ge 1$, while non-classical anti-bunched radiation has $g^{(2)} \left( 0 \right) < 1$. To see where the condition $g^{(2)} \left( 0 \right) > 1$ comes from, we look at the detector response to a classical wave with a time-dependent frequency and phase so that \eqref{strain3} becomes

\begin{equation}
h\left( t-z_a \right) = h_a (t)\:e^{ i\omega \left( t \right)t} e^{-i\phi \left( t \right)} ,
\label{strain4}
\end{equation}

\noindent where $h_a (t)$ is the strain amplitude at detector $a$ located at position $z_a$ and time $t$. Substituting \eqref{strain4} into \eqref{hCorrelation} the first order correlation becomes

\begin{equation}
g^{\left( 1 \right)} \left( \tau  \right) = \frac{{\left\langle {h_a \left( {t + \tau } \right) h_b (t)\:e^{-i\Delta \omega \left( t \right)t} e^{ i\Delta \phi \left( t \right)} e^{-i\omega \left( {t + \tau } \right)\tau } \:} \right\rangle }}{\sqrt {\left\langle { h_a \left( {t + \tau } \right)\:h_a \left( {t + \tau } \right)\:} \right\rangle }{\sqrt {\left\langle { h_{b} (t) h_{b} (t)\:} \right\rangle } } }.
\label{gw1correleation}
\end{equation}

\noindent For a single amplitude and frequency this first order correlation simplifies to \cite{Fox},

\begin{equation}
g^{\left( 1 \right)} \left( \tau  \right) = e^{-i\omega _0 \tau } \left\langle {e^{ i\Delta \phi \left( t \right)} } \right\rangle .
\label{gw1simple}
\end{equation}

\noindent For a single phase, with $\Delta \phi (t) =0$, one has $\left| {g^{\left( 1 \right)} \left( \tau  \right)} \right| = 1$ consistent with coherent radiation.

The second order correlation for the time varying amplitude is obtained by substituting \eqref{strain4} into \eqref{2Correlation} yielding

\begin{equation}
g^{\left( 2 \right)} \left( \tau  \right) = \frac{{\left\langle { h_a^2 (t)\: h_b^2 \left( {t + \tau } \right)\:} \right\rangle }}{{\left\langle {h_a^2 (t)} \right\rangle \left\langle {h_b ^2\left( {t + \tau } \right)} \right\rangle }} ,
\label{gw2correleation}
 \end{equation}

\noindent where $h_a^2 (t)$ and $h_b^2 (t)$ are the square amplitude \cite{Vannucci80,Facao11} of the strain in \eqref{strain4} at detectors $a$ and $b$. The square amplitudes from \eqref{gw2correleation} can be expanded \cite{Fox} at $\tau = 0$ as, $h_a^2 (t) = h_b^2 (t) = \left\langle {h^2} \right\rangle + \Delta h^2 \left( t \right)$, where $\left\langle {\Delta h^2 \left( t \right)} \right\rangle =0$, and thus \eqref{gw2correleation} becomes

\begin{equation}
g^{\left( 2 \right)} \left( 0  \right) = 1 + \frac{{\left\langle [\Delta h^2 \left( t \right)]^2 \right\rangle }}{{\left\langle {h^2} \right\rangle^2 }}~.
\label{gw3correleation}
\end{equation}

\noindent The second term on the RHS in \eqref{gw3correleation} is always  greater than zero ({\it i.e.} ${\left\langle [\Delta h^2 \left( t \right)]^2 \right\rangle }/{\left\langle {h^2} \right\rangle^2 } >0$) for non-coherent radiation, which then gives the condition $g^{\left( 2 \right)} \left( 0  \right) >1$. This relation is a general property of amplitude modulated signals \cite{Lebreton13} like those generated by CBC discussed in Section \ref{binary}.

%This demonstrates, in general $g^{\left( 2 \right)} \left( \tau  \right) \ne 1$, and that the gravitational radiation associated with the amplitude from \eqref{strain4} is not completely coherent.

\section{Intensity correlations for CBC} \label{binary}

In most applications of intensity correlations in optics the signals are periodic and the interval for time averaging is taken to be one period, $T$. This definition of the time average makes sense for periodic signals where the average is the same for each period. Since GW signals from CBC are transient the correlations in principle should be averaged over all time. However, averaging over all time is not possible using the conventional time averaging for second order correlations \eqref{correlationfunction} since the integrals are finite and the time interval is not. We must examine the method of time averaging in the correlations more carefully and in particular note that the time average must be taken over a specified time interval.

%\pagebreak

%-----------------------------Figure Start------------------------------
% encapsulated postscript
\begin{figure}[htp!]
 \centering
\includegraphics[width=100mm]{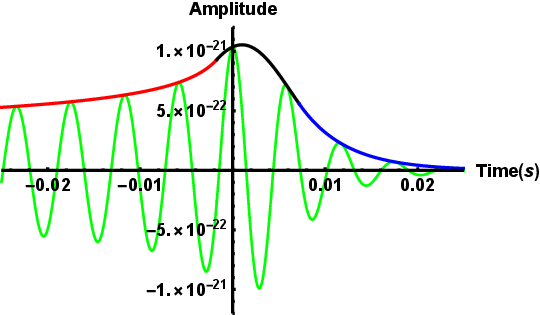}
 \caption{Toy model of the GW discovery signal including the magnitude of the strain amplitude $ \left| h \left( t \right) \right|$. The model was developed in Mathematica, solving for matching amplitudes and slopes for the inspiral/merger and merger/ringdown transitions. The inspiral, merger, and ringdown amplitudes are depicted in red, black, and blue respectively.}
\label{fig1}
\end{figure}
%-----------------------------Figure End--------------------------------

The expectation value is defined \cite{Mandel95} as $\left\langle {f(x)} \right\rangle = \int f(x) p(x)dx$ where $p(x)dx$ is the probability over the interval $dx$. On the other hand the time averages in the correlation functions are more broadly defined as noted by Glauber \cite{Glauber06}, ``The angular brackets $\left\langle{ ... }\right\rangle$ indicate that an average value is somehow taken, as we have noted.". Based on a suitable method for calculation of time averages the normalized intensity correlations \eqref{gw2correleation} are defined as,

\begin{equation}
\label{g2}
     g^{2}(\tau) =\frac{G^{(2)}(\tau)}{ G^{(1)}(t)G^{(1)}(\tau)}.
\end{equation}

\noindent If the signal is periodic then the average is the same for each period, $T$, and the integral time average takes the conventional form, $G^{(2)}(\tau)=1/T\int^{T/2}_{-T/2}I(t)I(t+\tau)dt$. Using this method of time averaging the normalized intensity correlations can be written in terms of the time interval $\Delta T = T_2 - T_1$ as,

\begin{equation}
\label{g2T}
     g^{2}(\tau) =\Delta T \cfrac{\int^{T_2}_{T_1}I(t)I(t+\tau)dt}{\int^{T_2}_{T_1}I(t)dt \int^{T_2}_{T_1}I(t+\tau)dt}.
\end{equation}

\noindent The time interval $\Delta T$ in the conventional scheme for time averaging does appear explicitly in the expression for the time average. This dependence is unavoidable with the conventional method of time average and suggest that other methods of time averaging might be worthy of consideration, e. g. ``intensity weighted averaging" \cite{Diels06}.

%-----------------------------Figure Start------------------------------
% encapsulated postscript
\begin{figure}[htp!]
 \centering
\includegraphics[width=100mm]{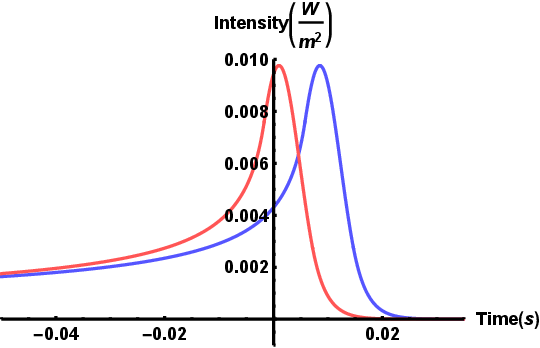}
 \caption{two-point intensities with lag time $\tau $ for spatially separated and concurrently operated independent detectors. The lag time $\tau$ is based on the geometry of the detectors and the orientation to the signal source. This lag time was approximately $\tau \approx 7 ~ \rm{ms} $ for the discovery signal detections \cite{Abbott}.}
\label{fig2}
\end{figure}
%-----------------------------Figure End--------------------------------

To further develop the gravito-optics of correlation functions for CBC and to compare to correlation functions in optics we will use a toy model of the GW discovery detection as shown in Fig. \ref{fig1}. To illustrate the calculation of intensity correlations for transient signals the model is developed using Mathematica and consist of an inspiral, merger, and ring-down. The inspiral is modeled by a realistic calculation \cite{Moreno17} using the parameters of the GW discovery \cite{Abbott}. The ring-down is modeled by an exponentially declining sinusoidal. The merger is modeled as a Gaussian connecting the amplitude and slope of the inspiral and the ring-down. The toy model illustrated in Fig. \ref{fig1} is well suited for intensity correlation calculations since the amplitude is easily extracted from the model waveform.

%\href{run:./ToyModel.nb}{Mathematica} 

%-----------------------------Figure Start------------------------------
% encapsulated postscript
\begin{figure}[htp!]
 \centering
\includegraphics[width=100mm]{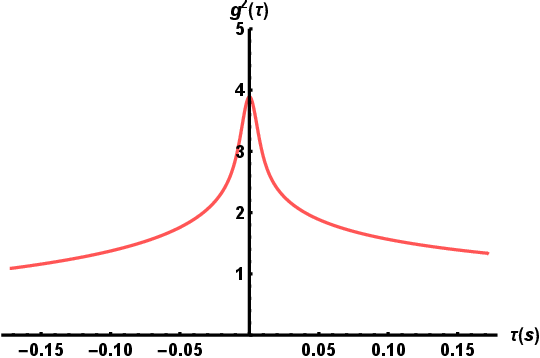}
 \caption{Intensity correlations, $g^{2}(\tau)$ as a function of lag time $\tau$, and taking averages over the time interval $T=1~\rm{s}$. The normalized intensity correlations can be considered a measure of the statistical correlation of two data sets. Higher order correlations would provide additional quantitative measures of statistical correlation, {\it e.g.} skewness and kurtosis \cite{Sun16}. }
\label{fig3}
\end{figure}
%-----------------------------Figure End--------------------------------

The intensity correlation functions \eqref{g2} are dependent on the time varying strain amplitude at two separate detectors. These intensities are calculated from our model in \eqref{g2T} with $T_2 = -T_1 =T/2$ where $T$ is the interval of the time averaging. For illustration the second order correlations between two detectors for different values of lag time $\tau$ are presented in Fig. \ref{fig2}. A lag time between detections of $\tau \approx 7 ~ \rm{ms} $ is assumed consistent with the discovery detection \cite{Abbott}. The normalized intensity correlations \eqref{g2T} for the two point detections are presented in Fig. \ref{fig3}. The baseline for the current two-point detections of GWs, see Fig. \ref{fig}, is on the order of one wavelength -- assuming a GW frequency of approximately $100 ~\rm{Hz}$ and a baseline of $3,000 ~\rm{km}$. For BH-BH mergers the coherence time \cite{Malvimat14} is $1/\Delta \nu \approx 1 ~ \rm{ms} - 10 ~ \rm{ms}$, the lag time $0 \le  \tau  \le 10 ~ \rm{ms}$, and detectable signal duration on the order of $100 ~ \rm{ms}$. These estimates and our toy model calculations, Fig. \ref{fig3}, show that although the existing GW observatories were never intended to be used as an HBT interferometer the concurrent operation of the observatories is nevertheless well suited for intensity interferometry.

% The details of these calculations are provided in Appendix \ref{Mathematica}. 

%\pagebreak

\section{The optical equivalence theorem} \label{equivalence}

The previous sections have described and developed many of the interesting similarities and contrast between applications of intensity correlations to optics and gravito-optics. Since both photons and gravitons are bosons one would expect that it should be equally possible to develop a theory of quantum gravito-optics comparable to the theory of quantum optics. There are two significant challenges to the development of a viable non-classical theory of gravito-optics that are not as serious in the classical theories. One is the unique problem that gravitons are the hypothetical quantization of gravitational radiation while photons are known quantizations for electromagnetic radiation. The second challenge is the optical equivalence theorem \cite{Unnikrishnan15,Sudarshan63,Mandel95,Ragy13} which demonstrates that classical and non-classical intensity correlations are formally equivalent for normal time ordering of quantized states and classical fields.

Where the optical equivalence theorem applies in electromagnetic radiation detections one must turn to well developed theories of quantum optics to provide a means of distinguishing between classical and non-classical processes. The challenge of distinguishing between classical and non-classical processes is described by Mandel and Wolf \cite{Mandel95}, ``Needless to say, as was emphasized by Glauber (1963b), the theorem does not imply the equivalence of quantum electrodynamics and classical optics." However, they go on to describe the formal challenge presented by the optical equivalence theorem,  ``Still, in those cases in which  $\phi \left( \upsilon \right) $ does behave like a classical probability density, there is indeed little difference between the evaluation of quantum expectations and the corresponding classical expectations." In many cases the distinction between classical and non-classical phenomena must be made through a good understanding of the experiment and the theoretical description of the processes studied. In the absence of a well established theory of quantum gravito-optics the formal equivalence of the correlations is a serious problem for distinguishing between classical and non-classical phenomena of gravitational radiation.

To begin the development of a theory of quantum gravito-optics we present the optical equivalence theorem \cite{Sudarshan63} in the form of gravitational fields and graviton creation/annihilation operators and follow up with a discussion of non-classicality criteria in Section \ref{multi-graviton}. Following \cite{Giovannini17} the nonclassical expectation values in terms of unweighted correlation functions to first order are,

\begin{equation}
\mathcal{G}^{(1)} \left(x_1,x_2 \right) = \left\langle { {{\hat h}^{\left(  -  \right)}} \left( x_1 \right) {{\hat h}^{\left(  +  \right)}} \left( x_2 \right) } \right\rangle .
\label{G1expectation}
\end{equation}

\noindent Similarly the classical unweighted correlation functions to first order are,
unweighted
\begin{equation}
{G}^{(1)}_{c} \left(x_1,x_2 \right) = {\left\langle { h^* \left( x_1 \right) h\left( x_2 \right)} \right\rangle } ,
\label{G1classicalx1x2}
\end{equation}

\noindent which can generally be written in terms of the lag time $\tau$,

\begin{equation}
{G}^{(1)}_{c} \left(x_1,x_2 \right) = {\left\langle { h^* \left( {t + \tau } \right) h\left( t \right)} \right\rangle } .
\label{G1classical}
\end{equation}

\noindent The unweighted second order nonclassical correlation functions are,

\begin{equation}
\mathcal{G}^{(2)} \left(x_1,x_2 \right) = \left\langle {{{\hat h}^{\left(  -  \right)}} \left( x_1 \right) {{\hat h}^{\left(  -  \right)}} \left( x_2 \right) {{\hat h}^{\left(  +  \right)}} \left( {x_2} \right) {{\hat h}^{\left(  +  \right)}} \left( x_1 \right)} \right\rangle ,
\label{G2expectation}
\end{equation}

\noindent and the classical second order correlation functions are,

\begin{equation}
{G}^{(2)}_{c} \left(x_1,x_2 \right) = {\left\langle {h^* \left( t \right)h^* \left( {t + \tau } \right)h\left( {t + \tau } \right)h\left( t \right)} \right\rangle }.
\label{G2expectation}
\end{equation}

\noindent According to the optical equivalence theorem with normal ordered or commutative quantum operators the classical and nonclassical correlation expectations are equivalent,

\begin{eqnarray}
        \label{OpticalEquivalence1}
        {G}^{(1)}_{c} \left(x_1,x_2 \right) =  \mathcal{G}^{(1)} \left(x_1,x_2 \right),\\
        \label{OpticalEquivalence2}
      {G}^{(2)}_{c} \left(x_1,x_2 \right) = \mathcal{G}^{(2)} \left(x_1,x_2 \right).
\end{eqnarray}

\noindent The only previous work that we know of in the development of the theory of quantum gravito-optics for inspirals \cite{Unnikrishnan15} treats the binary as a quantized potential in the Schr{\"o}dinger equation. This work also discusses the importance of the optical equivalence theorem in the development of quantum gravito-optics.

\section{Detector response} \label{exchange}

In this section we discuss the detector response \cite{Baym98,Baym99,Malvimat14,Wilczek15} when two or more detectors are operated simultaneously as an HBT interferometer. To this end we transition from classical fields to quantum mechanical creation and annihilation operators, $ h\left( t \right) \to \hat h \left( t \right)$. Following \cite{Giovannini19} we can split the quantum gravitational field into graviton annihilation operators, $\hat h^{\left( + \right)} \left( x  \right)$, and graviton creation operators,  $\hat h^{\left( - \right)} \left( x  \right)$, where $x$ is the space-time location at which the creation/annihilation operator is evaluated. Here we extend the work of references \cite{Baym98,Baym99,Giovannini11} from pions to gravitons. The demonstration of the detector enhancement starts with the probability of detecting a single graviton of energy $k$ with amplitude $h$ at detector $a$. This quantity is given by the expression

\begin{equation}
P^{a}_{h} \left(k\right)  = \int {dx_1dx_2} \mathcal{A}_k \left( {x_1,x_2} \right) \left\langle {{{\hat h}^{\left(  -  \right)}}\left( x_1 \right){{\hat h}^{\left(  +  \right)}}\left( {x_2} \right)} \right\rangle ~,
\label{enhanced1}
\end{equation}

\noindent where the notation for the ``response function''\footnote{Ayala {\it et al.} \cite{Baym99} describe this as a ``spectrometer function'', but the term ``response function'' is probably a better description for the GW detection process.}, $ \mathcal{A}_k$, has been adopted to support the introduction of tensor polarizations in Section  \ref{polarization}. Here $\left\langle {{{\hat h}^{\left(  -  \right)}}\left( x_1 \right){{\hat h}^{\left(  +  \right)}}\left( {x_2} \right)} \right\rangle$ is the single-graviton correlation function and $\mathcal{A}_k \left( {x_1,x_2} \right)$ is a response function for the detector $a$ over an interval $t_1-t_2$. Next one needs the probability for detection of two gravitons of energies $k$ and $k'$ at two different detectors $a$ and $b$, 

\begin{equation}
P^{ab}_{hh} \left(k,k'\right)  = \int {dX} \mathcal{A}_k \left( {x_2,x_3} \right)   \left\langle {{{\hat h}^{\left(  -  \right)}}\left( x_1 \right) {{\hat h}^{\left(  -  \right)}}\left( x_2 \right) {{\hat h}^{\left(  +  \right)}}\left( {x_3} \right) {{\hat h}^{\left(  +  \right)}}\left( x_4 \right)} \right\rangle  \mathcal{B}_{k'} \left( {x_1,x_4} \right),
\label{enhanced3}
\end{equation}

\noindent where to compact the notation we define $dX \equiv dx_1 dx_2 dx_3 dx_4$. The space-time coordinates $x_1, x_2, x_3$ and $x_4$ represent the time and position of the graviton creation or annihilation operators \cite{Baym99,Giovannini19,Glauber06} in the second order correlation $\left\langle {{{\hat h}^{\left(  -  \right)}}\left( x_1 \right) {{\hat h}^{\left(  -  \right)}}\left( x_2 \right) {{\hat h}^{\left(  +  \right)}}\left( {x_3} \right) {{\hat h}^{\left(  +  \right)}}\left( x_4 \right)} \right\rangle $ and in the two detector response functions $ \mathcal{A}_k \left( {x_2,x_3} \right) $ and $\mathcal{B}_{k'} \left( {x_1,x_4} \right) $. To practically make use of this condition one needs to be able to determine the response functions, like $\mathcal{A}_k$ and $\mathcal{B}_{k'}$ as well as determining the single-graviton and double graviton correlation functions like $\left\langle {{{\hat h}^{\left(  -  \right)}}\left( x_1 \right){{\hat h}^{\left(  +  \right)}}\left( {x_2} \right)} \right\rangle$ and $\left\langle {{{\hat h}^{\left(  -  \right)}}\left( x_1 \right) {{\hat h}^{\left(  -  \right)}}\left( x_2 \right) {{\hat h}^{\left(  +  \right)}}\left( {x_3} \right) {{\hat h}^{\left(  +  \right)}}\left( x_4 \right)} \right\rangle$. This has been done for pions, for example in \cite{Baym98, Baym99,GGLP}, and for photons, for example in \cite{Mandel95}.

The two problems in developing a theoretical framework for calculating, $P^{ab}_{hh} \left(k,k'\right)$, $P^{a}_{h} \left(k\right)$, and $P^{b}_{h} \left(k'\right)$, for GWs are determining the detector response functions ({\it e.g.} $\mathcal{A}_k$ and $\mathcal{B}_{k'}$) and the correlation functions ({\it e.g.} $\langle {{{\hat h}^{\left(  -  \right)}}\left( x_1 \right){{\hat h}^{\left(  +  \right)}}\left( {x_2} \right)} \rangle$). Assuming a classical detector response\footnote{See for example \cite{Cornish01,Thrane09,Jaranowski12,Romano17,Mukherjee19} where the classical response function is developed for GWs assuming that the detector responses are classical.}  in \eqref{enhanced1}  and \eqref{enhanced3} cannot be justified {\it a priori}. This is certainly not the case for the non-classical detector responses of Standard Model fields.  In particle physics the signal consists of particles ({\it i.e.} pions) and the detector response is the production of tracks through interactions of the particles with the detector fields \cite{Baym98}. In quantum optics the signal consist of photons and the detector response is due to electronic transitions in a photodetector \cite{Mandel95}. The characterization of the detector response for gravitons is more challenging because the quantum nature of the GW signal is not known. There has been some recent work on both the Newtonian \cite{Lieu17v2} and general relativistic \cite{Chen18} characterization of GW detectors in single graviton detection processes. The latter work in particular could prove useful in developing a theoretical framework for the representation of the single graviton response function in \eqref{enhanced1}. This could then be used to develop a multi-graviton detector response such as given in \eqref{enhanced3}.

As for the second problem, we are aware of only one theoretical work on the non-classical correlation functions associated with inspirals \cite{Unnikrishnan15}. There is however a well established theoretical framework for the non-classical correlation functions associated with primordial GWs \cite{Giovannini11,Giovannini17,Giovannini17b,Giovannini19a,Kanno18,Giovannini19,Kanno19}. Much of the established theory is sufficiently general that expanding this work to CBC should be much less challenging than the problem of the detector response functions.

\subsection{Polarization states and the detector response} \label{polarization}

The previous section describes the detector enhancement for concurrent detector operation but does not consider the graviton polarization states. We can extend this discussion to include polarization by assuming a plane wave with plus and cross polarizations. Here we will build on earlier work \cite{Giovannini19} introducing the polarization states in the correlation functions and extend this to include the detector response. Introducing the polarization into the correlation functions of the detection probability \eqref{enhanced3} leads to replacing the scalar response functions with associated tensor response functions,
\begin{equation}
P^{ab}_{hh} \left(k,k'\right)  = \int {dX} \mathcal{A}^{iji'j'}_k\left( {x_2} \right) \left\langle {{{\hat h}^{\left(  -  \right)}}_{lm} \left( x_1 \right) {{\hat h}^{\left(  -  \right)}}_{ij} \left( x_2 \right) {{\hat h}^{\left(  +  \right)}}_{i'j'} \left( {x_2} \right) {{\hat h}^{\left(  +  \right)}}_{l'm'} \left( x_1 \right)} \right\rangle   \mathcal{B}^{lml'm'}_{k'}\left( {x_1} \right).
\label{polarization1}
\end{equation}

\noindent We have also identified the time and position of detections at detector $a$ and detector $b$ with $x_2$ and $x_1$ respectively. Since the polarization response functions are diagonal these simplify to $\mathcal{A}^{iji'j'}_k\left( {x_2} \right) = \eta_{k} F^{A} F^{A'}$ and $   \mathcal{B}^{lml'm'}_{k'}\left( {x_1} \right)  = \eta_{k'} F^{B}F^{B'}$,
\begin{equation}
P^{ab}_{hh} \left(k,k'\right)  = \eta_{k} \eta_{k'}  \int {dX} F^{A} F^{A'}  F^{B}F^{B'} \left\langle {{{\hat h}^{\left(  -  \right)}}_{B} \left( x_1 \right) {{\hat h}^{\left(  -  \right)}}_{A} \left( x_2 \right) {{\hat h}^{\left(  +  \right)}}_{A'} \left( {x_2} \right) {{\hat h}^{\left(  +  \right)}}_{B'} \left( x_1 \right)} \right\rangle,
\label{polarization2}
\end{equation}
\noindent where $F^{A }$ is the polarization specific antenna response at ${x_2}$ and $\eta_{k}$ is the associated frequency dependent detection efficiency. The detection probability \eqref{polarization2} characterizes the interferometric GW detectors as intensity detectors as opposed to the conventional characterization as classical amplitude detectors.

Gravitational plane waves have two possible polarizations states, $A=+$ and $A=\times$. The response of a interferometric detector of arm length, $L$, can be written as \cite{Schutz09},
\begin{equation}
    \label{dll}
\frac{\delta L}{L} = F^{+} (\theta, \phi, \psi) h_{+} (t) + F^{\times} (\theta, \phi, \psi) h_{\times}  (t) ,
\end{equation}
where $h_{+, \times}$ are the two different polarization amplitudes and $F^{+, \times}$ are antenna response functions which depend on the orientation of the detectors with respect to the direction and polarization of the incoming GW. The angles $\theta$ and $\phi$ are the usual spherical polar angles. The $xy$ plane is the plane of the arms of the interferometer, $\phi$ is the azimuthal angle and $\theta$ is the polar angle between the $z$-axis which is perpendicular to the plane of the detector and the direction of propagation of the GW. The angle $\psi$ relates the orientation of the polarization tensors and the detector orientation in the sky plane \cite{Schutz09}. In terms of the relative angles between the detector orientation and the GW one can write down the antenna response functions as \cite{Schutz09},
\begin{eqnarray}
        \label{antenna}
        F^+ (\theta, \phi, \psi) &=& \frac{1}{2} (1 + \cos ^2 (\theta)) \cos (2 \phi) \cos (2 \psi) -\cos (\theta) \sin (2 \phi) \sin (2 \psi) ,\\
        \label{antenna1}
        F^\times (\theta, \phi, \psi) &=& \frac{1}{2} (1 + \cos ^2 (\theta)) \cos (2 \phi) \sin (2 \psi) + \cos (\theta) \sin (2 \phi) \cos (2 \psi)  .
\end{eqnarray}
One can see that the maximum value of $F^+$ and $F^\times$ is $1$. Further the current GW observatories are well separated on the Earth and the antenna response functions will be different at each detector since the angles, $\theta$, $\phi$, and $\psi$ will generally be different at each detector due in part to the curvature of the Earth. The overall effect is to reduce the detection probabilities like $P^{ab} _{hh} (k, k')$. The antenna response functions are also useful in determining the relation between the lag time, $\tau$, and the direction of the signal from the detectors \cite{Schutz11}.

\subsection{Detector efficiency and the detector response} \label{efficiency}

Following the characterization of photo-detection in optics \cite{Mandel95} we will represent the GW detector response in terms of a graviton detection efficiency, $ \eta_{k} = \frac{F_d}{F_g}$, the ratio of the effective graviton flux interaction with the detector and the incident gravitational radiation flux. An important distinction must be made here between our definition of $\eta$ and the more common use as the technical efficiency of the detector in responding  to the signal. The technical efficiency of a gravitational wave detector in responding physically to the signal is effectively one hundred percent. The use of the term efficiency used here is formally justified by the way $\eta$ appears in \eqref{polarization2}. The gravitational radiation flux is a function of the strain amplitude and frequency \cite{Schutz96}, $ F_g = \frac{c^3 h^2 \omega^2}{16 \pi G} $, which is comparable to the intensity in Section \ref{binary}. We can develop an idea of the magnitude of the graviton detection flux thru a ``back of the envelope'' calculation \cite{Lieu17v2}. Writing the equation of motion for the mirror displacement from the equilibrium position as, $ \ddot{\xi} = \frac{1}{2} \omega^2 h L e^{i \omega t} $, the average kinetic energy per cycle is then, $ E =  \frac{1}{2} M \dot{\xi}^2 = \frac{1}{8} M h^2 L^2 \omega^2 $, where $M$ is the mirror mass and $L$ the detector arm length.  This energy per cycle is associated with an effective flux of $ F_d =  \frac{1}{16 \pi} M h^2 \omega^3 $. Collecting terms the frequency dependent detector efficiency is, $ \eta_{k} = \frac{G}{c^3} M \omega $. If we assume a $50 ~ \rm{kg} $ mirror and angular frequency of $ 10^3 ~  \rm{\frac{rad}{s}} $ the detector efficiency is on the order of $ \eta \sim 10^{-31} $. This extremely low detector efficiency is not surprising considering the challenges that had to be overcome to extract the GW signal from detection noise.

Intensity correlations with low detector efficiencies are common in quantum optics and do not present any problem. For example in an application of intensity correlations in quantum optics with $\eta \sim 10^{-4}$ the lack of any affect was expressed as \cite{Koashi93}, ``It is worth noting that the low detection efficiency does not affect the measured value of $g_n^{(2)} $ since in Eq. (1) both numerator and the denominator are multiplied by $\eta^2$.". As stated earlier the relationship between $\eta$ in our discussion and technical efficiencies of optical detectors is formal justified here because it will cancel in the same way for normalized correlation functions. Our extremely low graviton detector efficiency of $ \eta \sim 10^{-31} $ has no affect on the value of the normalized intensity correlations since the terms cancel in the numerator and denominator. We will discuss the low detector particle number expectation associated with this low efficiency in the context of future work on quantum-gravito optics in Section \ref{multi-graviton}.

The fundamental physical principles applied to characterize the detection efficiency in our ``back of the envelope'' calculation are Newtonian. However, Pang and Chen \cite{Chen18} have shown that the formal Newtonian and general relativistic characterizations of the detectors differ by a gauge condition. Therefore, this order of magnitude calculation is a good representation of the actual detection efficiency of current GW detectors. Our characterization of the detector response has followed that of quantum optics using a phenomenological approximation of detection efficiency. It should be possible in future work to build on previous non-classical characterizations of laser interferometric GW detectors \cite{Meers88,Chen18} to describe the quantum mechanical response of the detectors. This would be consistent with the characterization of the detector response used in particle physics \cite{Baym98}.

\section{Quantum gravito-optics} \label{multi-graviton}

One of the most important contributions of intensity correlations to our understanding of optics is the associated development of quantum optics. A comparable development of quantum gravito-optics associated with correlations in gravito-optics is far more challenging due to the fact that gravitons are hypothetical. Since we do not know if there are non-classical phenomena in the detection of gravitational radiation from CBC, characterization of any such phenomena is ambiguous. To overcome this ambiguity the development of some meaningful measure or criteria for non-classicality in needed. There has been a good deal of recent theoretical work on developing some criteria for non-classicality for continuous GWs \cite{Giovannini11,Giovannini17,Giovannini17b,Giovannini19a,Kanno18,Giovannini19,Kanno19}. For insprirals this development seems to be limited to that of Unnikrishnan and Gillies \cite{Unnikrishnan15}.

Our principle goal in this paper is to provide a meaningful framework for the development of gravito-optics associated with GW detections from binary inspiral sources.  While the problem of quantum gravito-optics is beyond the scope of this paper having a good classical foundation is necessary to address that problem. Since the existence of the photon is not in doubt the criteria for non-classicality in optics is less important than for gravito-optics. The question of criteria or measures of non-classicality is sufficiently interesting that there has been significant theoretical work on this question. An incomplete but representative list of the potential measure of non-classicality that should be described by a fully developed theory of quantum gravito-optics is provided in Table \ref{table:graviton}.

%\begin{table}[t] % put at top of page if possible 
\begin{table}[ht]
\begin{tabular}{l|c|c}
 & Description & References \\
\hline

\begin{minipage}[t]{0.15\columnwidth}%
\vspace{5pt} 
 \begin{singlespace*}
\flushleft Graviton detectors  %
\end{singlespace*}
\vspace{10pt}
\end{minipage} & \begin{minipage}[t]{0.65\columnwidth}%
\vspace{5pt} 
 \begin{singlespace*}
\flushleft  The possibility of detecting single gravitons has been excluded based on fundamental physical arguments. %
\end{singlespace*}
\vspace{10pt}
\end{minipage} & \cite{Dyson13} \\
\hline

\begin{minipage}[t]{0.15\columnwidth}%
\vspace{5pt} 
 \begin{singlespace*}
\flushleft Squeezed states  %
\end{singlespace*}
\vspace{10pt}
\end{minipage} &  \begin{minipage}[t]{0.65\columnwidth}%
\vspace{5pt} 
 \begin{singlespace*}
\flushleft Sub-Poissonian intensity correlations are clearly distinguishable from classical correlations but the associated squeezed states of gravitons would not be produced by binary-inspirals. %
\end{singlespace*}
\vspace{10pt}
\end{minipage}  & \cite{Lovas97,Lovas01,Kanno18,Giovannini19,Kanno19,Grishchuk90} \\
\hline

\begin{minipage}[t]{0.15\columnwidth}%
\vspace{5pt} 
 \begin{singlespace*}
\flushleft Measurement probability  %
\end{singlespace*}
\vspace{10pt}
\end{minipage} & \begin{minipage}[t]{0.65\columnwidth}%
\vspace{5pt} 
 \begin{singlespace*}
\flushleft The detection probability is factored into direct and exchange terms where the exchange term can be associated with multiparticle detection processes described in Sections \ref{exchange} and  \ref{polarization}. %
\end{singlespace*}
\vspace{10pt}
\end{minipage} & \cite{Baym98,Baym99} \\
\hline

\begin{minipage}[t]{0.15\columnwidth}%
\vspace{5pt} 
 \begin{singlespace*}
\flushleft Many detectors  %
\end{singlespace*}
\vspace{10pt}
\end{minipage} & \begin{minipage}[t]{0.65\columnwidth}%
\vspace{5pt} 
 \begin{singlespace*}
 \renewcommand{\thempfootnote}{\arabic{mpfootnote}}
\flushleft  Intensity correlations for three or more detectors potentially provide information on phase, improvements in image reconstruction,  qualitative phase information\tablefootnote[4]{ This is an example of the ``phase problem'' and associated methods of phase retrieval \cite{Goldberger63}.}, and associations between classical field intensities and non-classical correlations. %
\end{singlespace*}
\vspace{10pt}
\end{minipage} & \cite{Malvimat14} \\
\hline

\begin{minipage}[t]{0.15\columnwidth}%
\vspace{5pt} 
 \begin{singlespace*}
\flushleft Quantum discord   %
\end{singlespace*}
\vspace{10pt}
\end{minipage} &  \begin{minipage}[t]{0.65\columnwidth}%
\vspace{5pt} 
 \begin{singlespace*}
\flushleft A measure of the non-classical correlations of the subsystems of a quantum system. %
\end{singlespace*}
\vspace{10pt}
\end{minipage}  & \cite{Kanno16,Kanno18,Hunt19,Kumar20,Ragy13,Modi10,Nambu14} \\
\hline

\begin{minipage}[t]{0.15\columnwidth}%
\vspace{5pt} 
 \begin{singlespace*}
\flushleft Entanglement entropy    %
\end{singlespace*}
\vspace{10pt}
\end{minipage} &  \begin{minipage}[t]{0.65\columnwidth}%
\vspace{5pt} 
 \begin{singlespace*}
\flushleft A measure of the Von Neumann entropy of entanglement of a two part composite system. %
\end{singlespace*}
\vspace{10pt}
\end{minipage}  & \cite{Kanno17,Kanno18,Benedetti20} \\
\hline

\begin{minipage}[t]{0.15\columnwidth}%
\vspace{5pt} 
 \begin{singlespace*}
\flushleft Entanglement negativity   %
\end{singlespace*}
\vspace{10pt}
\end{minipage} &  \begin{minipage}[t]{0.65\columnwidth}%
\vspace{5pt} 
 \begin{singlespace*}
\flushleft A measure of the violation of the separability criteria for the density matrix of mixed quantum states. %
\end{singlespace*}
\vspace{10pt}
\end{minipage}  & \cite{Kanno18,Badhani21} \\
\hline

\begin{minipage}[t]{0.15\columnwidth}%
\vspace{5pt} 
 \begin{singlespace*}
\flushleft R{\'e}nyi entropy   %
\end{singlespace*}
\vspace{10pt}
\end{minipage} &  \begin{minipage}[t]{0.65\columnwidth}%
\vspace{5pt} 
 \begin{singlespace*}
\flushleft Classical and non-classical correlations for low intensity sources are defined separately. %
\end{singlespace*}
\vspace{10pt}
\end{minipage}  & \cite{Ragy13} \\
\hline

\begin{minipage}[t]{0.15\columnwidth}%
\vspace{5pt} 
 \begin{singlespace*}
\flushleft Capacity of  entanglement  %
\end{singlespace*}
\vspace{10pt}
\end{minipage} &  \begin{minipage}[t]{0.65\columnwidth}%
\vspace{5pt} 
 \begin{singlespace*}
\flushleft A measure the width of the eigenvalue distribution or ``loosely thought of as the variance in the entanglement entropy".
\end{singlespace*}
\vspace{10pt}
\end{minipage}  & \cite{Boer19} \\
\hline

\begin{minipage}[t]{0.15\columnwidth}%
\vspace{5pt} 
 \begin{singlespace*}
\flushleft Bell's inequality  %
\end{singlespace*}
\vspace{10pt}
\end{minipage} &  \begin{minipage}[t]{0.65\columnwidth}%
\vspace{5pt} 
 \begin{singlespace*}
\flushleft Bell inequalities provide criteria for distinguishing between classical and quantum mechanical descriptions of detection processes. %
\end{singlespace*}
\vspace{10pt}
\end{minipage}  & \cite{Kanno18,Reid86,Choudhury17}  \\
\hline

\begin{minipage}[t]{0.15\columnwidth}%
\vspace{5pt} 
 \begin{singlespace*}
\flushleft Weak measurement  %
\end{singlespace*}
\vspace{10pt}
\end{minipage} &  \begin{minipage}[t]{0.65\columnwidth}%
\vspace{5pt} 
 \begin{singlespace*}
\flushleft Weak quantum measurements minimally disrupt a system (see Subsection \ref{efficiency}) while providing meaningful information about the observables of the system.  %
\end{singlespace*}
\vspace{10pt}
\end{minipage}  & \cite{Aharonov02,Carmichael04,Ragy13,Lieu17v2,Chen18,Bernardo14} \\
\hline

\begin{minipage}[t]{0.15\columnwidth}%
\vspace{5pt} 
 \begin{singlespace*}
\flushleft P-non classicality  %
\end{singlespace*}
\vspace{10pt}
\end{minipage} &  \begin{minipage}[t]{0.65\columnwidth}%
\vspace{5pt} 
 \begin{singlespace*}
\flushleft Glauber-Sudarshan P functions are not described as probability distributions for non-classical processes.   %
\end{singlespace*}
\vspace{10pt}
\end{minipage}  & \cite{Sudarshan63,Kiesel08,Ragy13,Bernardo14,Innocenti22} \\

\end{tabular}
\caption{Potential theoretical criteria for gravito-optical distinction of classical and non-classical processes in the detection of GWs.}
\label{table:graviton}
\end{table}

Not included in Table \ref{table:graviton} are potential theoretical criteria specifically based on first order correlations. Classically gravitational wave detectors respond to the signal amplitude which could be associated with both first and second order correlations. Optical detectors generally respond to signal energy which is naturally associated with second order correlations. However, second order correlations do not include the same information as first order correlations, {\it e.g.} phase information. It would be interesting to consider the potential theoretical criteria for both classical and non-classical gravitational wave detection processes for {\it two-point} and first order correlations. These studies are not included in Table \ref{table:graviton} because we are not aware of any examples of this line of research in the literature.

Of the theoretical measures of non-classicality described in Table \ref{table:graviton} that of R{\'e}nyi entropy \cite{Ragy13} is sufficiently well developed for a cursory comparison to the transient signals from binary inspiral sources.  The previous study is for photons and a great deal of further work is required to determine how well the measure can be applied to gravitons. Since both photons and gravitons are bosons and for the purpose of discussion we will assume that similar R{\'e}nyi entropy \cite{Ragy13} relations hold for non-classical contributions to two-point GW detections. The associated graviton-number expectation \cite{Fox,Lieu17v2} associated with the discovery and following Section \ref{efficiency} is $\langle \overline{n} \rangle \sim N = \frac{ F_d }{\hbar \omega} L^2 T \sim 10^{3} $ assuming a period of $10 ~ \rm{ms}$. This is at least one order of magnitude too great for significant non-classical contributions to the detection. For example we see in Figure 2 (c) of \cite{Ragy13}  that there is little likelihood of discernable non-classical contributions for $\langle \overline{n} \rangle > 100 $ using R{\'e}nyi entropy as a measure. Future GW detectors are expected to have several orders of magnitude greater sensitivity and are much more likely to exhibit non-classicality provided that non-classical gravitational phenomena exist. The average particle number $\langle \overline{n} \rangle$ is proportional to the intensity of the signal and not the amplitude. This relation is intrinsic to the non-classicality of gravitational radiation and intensity or second order correlations. As noted earlier, there might be interesting classical advantages to studies of first order correlations for gravitational radiation detection. However, these potential advantages would not hold equally for studies of non-classicality.

\section{Summary and conclusions}

The existing network of GW observatories have been operating serendipitously as an HBT interferometer since the first observed binary inspiral event. This makes it possible to consider intensity correlations and the gravito-optics of signal detections from CBC.  There have been earlier proposals for applying HBT interferometry to GWs, but these proposals have focused on periodic GW signals that are of cosmological origin \cite{Giovannini11,Giovannini17,Giovannini17b,Giovannini19a,Kanno18,Giovannini19,Kanno19} and are not currently detectable. Our focus here is on the intensity interferometry of transient signals from CBC that are currently detectable.

The importance of HBT interferometry and intensity correlations in the detection processes for both optics and particle physics has been well established over the past 60 years, beginning with the early work of Hanbury Brown and Twiss \cite{Brown56} and soon after by Glauber \cite{Glauber63,Glauber63_1} and Sudarshan \cite{Sudarshan63}. The existing literature on HBT interferometry for GW detection processes is far more recent and has only been developed for persistent signals associated with primordial GWs. We have focused on mitigating many of the difficulties of application of intensity correlations to transient signals from CBC. There are a great many fundamental problems that must be resolved in order to fully realize the potential significance of non-classical processes in the detection of transient signals associated with CBC. There is recent work \cite{Lieu17v2,Chen18} on the characterization of the GW detectors as quantum detectors, however this work is only preliminary. These works also focus on isolated single detectors while we have addressed two-point correlations between isolated and spatially separated detectors.  As we have shown these non-classical processes are not likely to contribute to existing detection processes but might be meaningful for third generation detectors.

We have taken some of the first steps in the development of a meaningful theory of gravito-optics for the better characterization and understanding of detection processes for CBC. As discussed in Section \ref{multi-graviton} there is more research needed for the development of a theory of quantum gravito-optics to better understand the non-classical nature of gravitational radiation. Nevertheless, the rough estimates in this paper demonstrate the promise of HBT interferometry and intensity correlations for GW detections from CBC.

%\appendix

%\section{Mathematica code for the toy model}  \label{Mathematica}

\end{document}